\documentclass[10pt]{revtex4}
\usepackage{graphicx} 
\usepackage{color}
\usepackage{babel}
\usepackage{fancybox}
\usepackage{calc}
\usepackage{amsmath}
\usepackage{amssymb}
\usepackage[a4paper]{geometry}
\usepackage{comment}
\usepackage[unicode=true,pdfusetitle,
 bookmarks=true,bookmarksnumbered=false,bookmarksopen=false,
 breaklinks=true,pdfborder={0 0 0},pdfborderstyle={},backref=false,colorlinks=true]
 {hyperref}

\begin{document}

\title{\Large  Reduced-order Bopp-Podolsky model on the null-plane}

\author{Mario C. Bertin}
\affiliation {São Paulo State University (Unesp), Institute
for Theoretical Physics (IFT), São Paulo, SP, Brazil.\\ (On leave of absence from Universidade Federal da Bahia (UFBA), Physics Institute, Salvador, Bahia, Brazil.)\\mbertin@ufba.br}

\author{Ronaldo Thibes}
\affiliation{Departamento de Ci\^encias Exatas e Naturais,
Universidade Estadual do Sudoeste da Bahia,
 45700-000, Itapetinga BA, Brazil.\\thibes@uesb.edu.br}

\begin{abstract}
We consider the null-plane dynamics for a reduced-order version of the higher-derivatives Bopp-Podolsky generalized electrodynamics model.  By introducing an auxiliary vector field, we achieve a simpler equivalent version with lower derivatives.  The massive and massless modes for the Podolsky gauge field get split into two sectors.  We describe the model in terms of light-front coordinates and, by choosing $x^+$ as a natural evolution parameter, proceed to its null-plane dynamics analysis obtaining the whole constraints structure and canonical field equations.  After a convenient constraints redefinition, we calculate the Dirac brackets corresponding to the second-class sector.  Gauge invariance is preserved and, after elimination of second-class constraints, we obtain a consistent Abelian first-class theory for the reduced-order Bopp-Podolsky model. 
\end{abstract}

\maketitle


\section{Introduction}
Since the closing of the last century, with the rescuing works of Pimentel, Barcelos-Neto, Galv\~ao and Natividade \cite{Galvao:1986yq, BarcelosNeto:1991dp}, we have been witnessing renewed interest in the long-lasting higher derivative Bopp-Podolsky (BP) model \cite{Bopp, Podolsky:1942zz, Podolsky:1944zz, Podolsky:1948}.  Recently revisited through the eyes of modern quantum field theory \cite{Bufalo:2013joa,
Bufalo:2014jra, Bufalo:2015koa, Mishra:2018hxc,
Ji:2019phv} and mathematical physics \cite{Lazar:2019svx, Chen:2021scz, Liu:2023aws, Missaoui}, BP electrodynamics has been more and more notably giving rise to new possible interpretations and interesting enchanting applications \cite{Bonin:2009je, Barone:2016pyy, Cuzinatto:2017srn, Bonin:2022tmg, Frizo:2022jyz, Raza:2024frc, deMelo:2024dxu,   Ferreira:2024qkd, Ferreira:2019lpu, Lisboa-Santos:2023pwc}. In fact, the BP model has been recatching the attention of the theoretical physics community due to its versatility as a building block for gauge-invariant massive theories and its many intriguing open issues related to the quantization of higher-derivative models.  Originally designed to tame infinities in the early era of quantum physics development, it can be related to the more modern regularization schemes of quantum field theory \cite{Ji:2019phv}. It can certainly provide an important ingredient for our quest towards the current major open problems in physics related to yet unexplained observational data in an accelerated expanding universe.  

Important subtleties of the BP model have been discussed in the late reference \cite{Ji:2019phv} where, besides being directly related to Pauli-Villars regularization, a thorough analysis of static density charge configurations has been worked out, and important natural higher-order gauge-fixing functions have been proposed for both covariant and light-front approaches. Those specific gauge-fixing functions have been shown to lead to simpler, neater expressions for the field propagators, a key fact for perturbative calculations.  BP model's main feature amounts to describing a higher-order derivative generalized electrodynamics, still linear in the fields, providing mass for the gauge field in a gauge invariant way in terms of a nontrivial mass parameter $a$. A couple of years ago, in \cite{Bonin:2022tmg}, Bonin and Pimentel showed how BP's generalized electrodynamics might arise from an Abelian Higgs model containing a nonminimally coupled scalar field for a certain regime, thus suggesting that the Podolsky mass parameter could emerge from a specific Higgs mechanism.  Also, the Aharonov-Bohm effect has been discussed in the context of the BP model.  The authors of \cite{deMelo:2024dxu} have found differences between Maxwell and BP electrodynamics regarding phase shift predictions, proposing experimental bounds to a possible physical mass for the photon.  The role played by BP electrodynamics in a possibly Lorentz symmetry violating scenario has been investigated in reference \cite{Ferreira:2024qkd}, where the authors have also considered contributions from a Carroll-Field-Jackiw term.

Concerning gravity and cosmological aspects, it is worth mentioning the role played by the BP model as a novel ingredient for trending works such as \cite{Cuzinatto:2017srn, Frizo:2022jyz, Raza:2024frc}.  In \cite{Cuzinatto:2017srn}, the authors studied black holes in the presence of
matter described by BP electrodynamics.  The BP model is generalized to curved space-time and coupled to Einstein gravity, defining an Einstein-Podolsky action.  An explicit viable wormhole solution was then found in \cite{Frizo:2022jyz} and recently generalized to the rotational case in \cite{Raza:2024frc}.

In this scenario, with so many new results appearing in a higher-derivative BP electrodynamics context, we draw our attention to the 2017 BJP article \cite{Thibes:2016ivt} containing a proposal for an alternative reduced-order version for the BP model.  As shown in \cite{Thibes:2016ivt}, by means of introducing an auxiliary companion vector field $B_\mu$ for Podolsky's, not only can a neat derivative order reduction with respect to the original BP proposal occur, but also a nice physical interpretation naturally shows up.  The massive and massless modes for the Podolsky gauge field split into two distinct parts and accommodated into different sectors of the theory, respecting the number of physical degrees of freedom.  The equivalence of approaches at the quantum level has been demonstrated in \cite{Thibes:2016ivt}.  Despite being further discussed in subsequent papers \cite{Ji:2019phv, Nogueira:2018jdm, Oliveira:2020xwj, Dai:2020qpc, ErrastiDiez:2023gme}, we feel that much more ought to be properly explored in the reduced-order BP proposal \cite{Thibes:2016ivt}, among which figures its null-plane analysis -- not yet carried out in the literature.  The filling of that gap constitutes precisely the main goal of this current letter.

The analysis of canonical structures on the null-plane shows some important advances compared to the usual instant-form dynamics. First, the stability group of the Poincaré group in the light-cone decomposition, which is the subgroup relating field configurations in a constant $x^+\equiv x^0 + x^1$  hyper-surface, has one more generator compared to the usual instant-form dynamics  \cite{Bakker, Ji:2018pic, Ma:2021yqx}. Also, the algebra of these generators is considerably simplified, since a boost becomes a simple scale transformation. As first pointed out by Dirac \cite{Dirac 1},  a Hamiltonian theory on light-cone coordinates presents extra second-class constraints, reducing the number of evident degrees of freedom. As a result, canonical quantization often results in excitation-free vacuum states and a better convergence of Feynman diagrams \cite{Casana:2010zz, Srivastava, Brodsky, Heinzl, Acevedo:2022qqs, Acevedo:2022chi}. Canonical analysis of BP electrodynamics on the null-plane may be found in \cite{Bertin:2009gs, Bertin:2017orz}.

For the reader's convenience, we have organized our work as follows.  In Sec II below, we characterize the reduced-order version for the BP generalized electrodynamics model and discuss some of its features passing through the field equations and energy-momentum tensor.  We introduce our notation and conventions for the null-plane in Sec III.  Proceeding in Sec IV, we apply the Dirac-Bergmann algorithm on the null-plane, obtaining its constraint structure, and follow to Sec V with their corresponding classification and consistency analysis.  In Sec VI, we compute the Dirac brackets corresponding to the second-class constraints sector of the theory and then come to the canonical field equations in Sec VII.  We end in Sec VIII with our final considerations.

\section{The Reduced-Order Bopp-Podolsky Model}
The reduced-order model for BP electromagnetic theory we shall analyze here is defined by the action \cite{Thibes:2016ivt}
\begin{equation}
S=\int_{\Omega}d\omega\left(-\frac{1}{4}F_{\mu\nu}F^{\mu\nu}-\frac{a^{2}}{2}B_{\mu}B^{\mu}+\frac{a^{2}}{2}G_{\mu\nu}F^{\mu\nu}\right),\label{eq:01}
\end{equation}
where $\Omega$ is a definite, but not specified, volume of the flat space-time, with $d\omega$ denoting its volume element, and $a$ the characteristic Podoslky's inverse mass parameter. Here, $F_{\mu\nu}\equiv\partial_{\mu}A_{\nu}-\partial_{\nu}A_{\mu}$
are the components of Maxwell's tensor with $A_{\mu}$ as the usual electromagnetic
field and $B_{\mu}$, responsible for the higher-derivatives order reduction, plays the role of an auxiliary vector field with $G_{\mu\nu}\equiv\partial_{\mu}B_{\nu}-\partial_{\nu}B_{\mu}$  standing for its corresponding field-strength tensor. Note
that the Lagrangian function in (\ref{eq:01}), unlike the full Lagrangian
of BP's original theory, has only first-order
derivatives of the fundamental fields.  The equivalence of (\ref{eq:01}) to the original higher-derivative descriptions of Bopp and Podolsky models \cite{Bopp, Podolsky:1942zz} in both classical and quantum levels has been fully established in references
\cite{Thibes:2016ivt, Oliveira:2020xwj}.

The field equations of the theory, obtained by imposing stationarity of (\ref{eq:01}) with respect to fundamental field variations, are given by\begin{subequations}\label{eq:02}
\begin{gather}
\partial_{\mu}F^{\mu\nu}-a^{2}\partial_{\mu}G^{\mu\nu}=0,\label{eq:02a}\\
a^{2}\left(\partial_{\mu}F^{\mu\nu}+B^{\nu}\right)=0,\label{eq:02b}
\end{gather}
\end{subequations}from which follows, for $a^{2}\neq0$,
\begin{equation}
B^{\mu}=a^{2}\partial_{\nu}G^{\mu\nu}.\label{eq:03}
\end{equation}
Eq. (\ref{eq:03}) then implies $\partial_{\mu}B^{\mu}=0$, and
\begin{equation}
\left(1+a^{2}\square\right)B^{\mu}=0,\label{eq:04}
\end{equation}
where $\square$ is the d'Alembertian operator. Moreover, eq. (\ref{eq:02b})
results in
\begin{gather}
\left(1+a^{2}\square\right)\partial_{\nu}F^{\nu\mu}=\left(1+a^{2}\square\right)\left(\delta_{\nu}^{\mu}\square-\partial^{\mu}\partial_{\nu}\right)A^{\nu}=0,\label{eq:05}
\end{gather}
reproducing the field equations of BP's theory.  The canonical energy-momentum tensor density can be obtained from (\ref{eq:01}) as
\begin{equation}\label{CEMT}
H_{\thinspace\thinspace\thinspace\nu}^{\mu} = \left(a^{2}G^{\mu\gamma}-F^{\mu\gamma}\right)\partial_{\nu}A_{\gamma}+a^{2}F^{\mu\gamma}\partial_{\nu}B_{\gamma}-\delta_{\nu}^{\mu}{\cal L}
\,,
\end{equation}
with $\cal L$ denoting the integrand of (\ref{eq:01}), and can be recast into the corresponding gauge-invariant symmetric form
\begin{eqnarray}
T^{\mu\nu}&=& F^\mu_{\;\;\gamma} F^{\gamma\nu}
+{1\over4}\eta^{\mu\nu}F^{\gamma\lambda}F_{\gamma\lambda}+{a^2\over2}\eta^{\mu\nu}\left(B^\gamma B_\gamma - G_{\gamma\lambda}F^{\gamma\lambda}\right)
\nonumber\\&&
+a^2F^\mu_{\;\;\gamma}G^{\nu\gamma}+a^2F^\nu_{\;\;\gamma}G^{\mu\gamma} - a^2 B^\mu B^\nu
\,.
\end{eqnarray}
It is interesting to see that in the reduced-order form, the well-known massive and massless modes for the Podolsky gauge field decouple in a precise way such that $B_\mu$ acquires the full Podolsky mass while $A_\mu$ becomes massless.  Gauge invariance is naturally preserved.  The canonical structure of the reduced-order BP model has been discussed in references \cite{Thibes:2016ivt, Oliveira:2020xwj} in terms of instant-form evolution.  However, its null-plane dynamics has not so far been addressed in the literature, an important gap we shall fill here, from next section on.

\section{The null-plane dynamics}
The null-plane dynamics is characterized by the choice of the null-plane coordinate $\tau\equiv x^{+}\equiv\frac{1}{\sqrt{2}}\left( x^0 + x^3 \right)$ 
as the evolution parameter of the considered theory. The dynamical
evolution of distributions of fields calculated in characteristic
surfaces of constant $\tau$ is evaluated by writing the field equation
in an appropriate coordinate system, the null-plane coordinates $\left(x^{+},x^{-},x^{1},x^{2}\right)$,
obtained from the usual space-time coordinates $\left(x^{0},x^{1},x^{2},x^{3}\right)$
by the transformation matrix 
\[
\frac{1}{\sqrt{2}}\left(\begin{array}{cccc}
1 & 0 & 0 & 1\\
1 & 0 & 0 & -1\\
0 & \sqrt{2} & 0 & 0\\
0 & 0 & \sqrt{2} & 0
\end{array}\right).
\]
The components of Minkowski's metric are now given by
\begin{equation}
\left({\hat{\eta}}_{\mu\nu}\right)\equiv\left(\begin{array}{cccc}
0 & 1 & 0 & 0\\
1 & 0 & 0 & 0\\
0 & 0 & -1 & 0\\
0 & 0 & 0 & -1
\end{array}\right), \label{eq:08}
\end{equation}
with $\mu,\nu=+,-,1,2$,
which should be used to raise and lower null-plane indexes.

It is always important to stress that the transformation to the null-plane
coordinates is a simple coordinate transformation, but it is not a
reference frame transformation, as it becomes clear by (\ref{eq:08}) showing the metric non-invariance \cite{Thibes:2022ssf}.
Furthermore, the derivative operators can be written as
\begin{equation}
\partial_{\mu}\equiv\left(\partial_{+},\partial_{-},\partial_{i}\right),\thinspace\thinspace\thinspace\thinspace\partial^{\mu}\equiv\left(\partial_{-},\partial_{+},-\partial_{i}\right),\thinspace\thinspace\thinspace\thinspace\thinspace\thinspace\thinspace\thinspace i=1,2,\label{eq:09}
\end{equation}
while the d'Alembertian operator takes the form
\begin{equation}
\square\equiv\partial_{\mu}\partial^{\mu}=2\partial_{-}\partial_{+}-\nabla^{2},\thinspace\thinspace\thinspace\thinspace\thinspace\thinspace\thinspace\nabla^{2}\equiv\partial_{i}\partial_{i}.\label{eq:10}
\end{equation}
Hence, field differential equations have their order reduced concerning the evolution parameter
derivatives: (\ref{eq:05}) represents now an up to third-order differential equation in $\partial_{+}$,
while (\ref{eq:04}) is first-order in $\partial_{+}$. This fact, however, does
not alter the initial value data required for unique solutions, since
there are actually two characteristic surfaces of constant $x^{+}$.
Eq. (\ref{eq:04}) requires an initial condition for $B^{\mu}$ in
a constant $x^{+}$ surface, but also three boundary
conditions in constant $x^{-}$ and $x^{i}$ surfaces. Eq. (\ref{eq:05})
requires a total of 16 initial/boundary data. More details on null-plane dynamics
can be seen in \cite{Dirac 1,Bakker} as well as in the recent applications \cite{Casana:2010zz, Bertin:2009gs, Bertin:2017orz, Acevedo:2022qqs, Acevedo:2022chi}.

\section{Primary and secondary constraints}
For convenience, we introduce the notation $\bar{A}\equiv\partial_{+}A$ for a corresponding given observable
$A$. On null-plane coordinates, the Hamiltonian function $H$ can be found
from the canonical energy-momentum tensor density (\ref{CEMT}) as its double projection over
the vector $\left(u^{\mu}\right)\equiv\left(1,0,0,0\right)$:

\begin{eqnarray}
H & = & \int_{\Sigma}d\sigma u_{\mu}H_{\thinspace\thinspace\nu}^{\mu}u^{\nu}=\int_{\Sigma}d\sigma H_{\thinspace\thinspace+}^{+}\nonumber \\
 & = & \int_{\Sigma}d\sigma\left[\left(F^{\mu+}-a^{2}G^{\mu+}\right)\bar{A}_{\mu}+a^{2}F^{+\mu}\bar{B}_{\mu}-{\cal L}\right],\label{eq:13}
\end{eqnarray}
where $\Sigma$ is a characteristic surface of constant $\tau\equiv x^{+}$
with volume element $d\sigma=dx^{-}dx^{1}dx^{2}$. Therefore, the
canonical variables of the theory are the fields $A_{\mu}$ and $B_{\mu}$
conjugated respectively to the momenta\begin{subequations}\label{eq:14}
\begin{gather}
p^{\mu}\equiv F^{\mu+}-a^{2}G^{\mu+},\label{eq:14a}\\
\pi^{\mu}\equiv a^{2}F^{+\mu}.\label{eq:14b}
\end{gather}
\end{subequations}The definitions (\ref{eq:14}) result in the primary
constraints\begin{subequations}\label{eq:15}
\begin{gather}
\phi_{1}\equiv p^{+}\approx0,\\
\phi_{2}^{i}\equiv p^{i}-F_{-i}+a^{2}G_{-i}\approx0,\\
\phi_{3}\equiv\pi^{+}\approx0,\\
\phi_{4}^{i}\equiv\pi^{i}+a^{2}F_{-i}\approx0,
\end{gather}
\end{subequations}and the relations

\begin{subequations}\label{eq:16}
\begin{gather}
\bar{A}_{-}=\partial_{-}A_{+}-\frac{1}{a^{2}}\pi^{-},\label{eq:16a}\\
\bar{B}_{-}=\partial_{-}B_{+}-\frac{1}{a^{2}}\left(p^{-}+\frac{1}{a^{2}}\pi^{-}\right).\label{eq:16b}
\end{gather}
\end{subequations}

We find the canonical Hamiltonian substituting (\ref{eq:15}) and
(\ref{eq:16}) in (\ref{eq:13}) as

\begin{eqnarray}
H_{c} & = & \int_{\Sigma}d\sigma\left[-\frac{1}{2}\frac{1}{a^{4}}\left(\pi^{-}\right)^{2}-\frac{1}{a^{2}}p^{-}\pi^{-}+\left(p^{-}\partial_{-}+p^{i}\partial_{i}\right)A_{+}+\left(\pi^{-}\partial_{-}+\pi^{i}\partial_{i}+a^{2}B_{-}\right)B_{+}\right.\nonumber \\
 &  & \left.+\frac{1}{4}F_{ij}F^{ij}-\frac{a^{2}}{2}G_{ij}F^{ij}+\frac{a^{2}}{2}B_{i}B^{i}\right]\label{eq:17}
 \,,
\end{eqnarray}
from which the primary Hamiltonian can be written as
\begin{equation}
H_{P}\equiv H_{c}+\int_{\Sigma}d\sigma u^{a}\left(x\right)\phi_{a}\left(x\right)
\,,
\label{eq:18}
\end{equation}
with a repeated index sum convention in $a=1,2^{(i)},3,4^{(i)}$ to account for the six primary constraints (\ref{eq:15}) and $u^a$ denoting the corresponding Lagrange multiplier functions.
Using the primary Hamiltonian as the generator for the $\tau$ evolution,
we next impose Dirac's consistency conditions. For any primary
constraint $\phi_{a}$, consistency requires $\partial_{+}\phi_{a}\left(x\right)=\left\{ \phi_{a}\left(x\right),H_{P}\right\} \approx0$,
where $\left\{ \bullet,\bullet\right\} $ is the Poisson bracket (PB)
considering the phase-space variables $\left(A_{\mu},p^{\mu},B_{\mu},\pi^{\mu}\right)$.
The fundamental PBs of the theory are given by
\begin{equation}
\left\{ A_{\mu}\left(x\right),p^{\nu}\left(y\right)\right\} =\left\{ B_{\mu}\left(x\right),\pi^{\nu}\left(y\right)\right\} =\delta_{\mu}^{\nu}\delta^{3}\left(x-y\right),\label{eq:19}
\end{equation}
where $\delta^{3}\left(x-y\right)\equiv\delta\left(x^{-}-y^{-}\right)\delta\left(x^{1}-y^{1}\right)\delta\left(x^{2}-y^{2}\right)$
is the three-dimensional Dirac delta in null-plane coordinates.

In order to unravel the full set of constraints with their corresponding algebra and gauge properties, we follow the well-known Dirac-Bergmann algorithm \cite{Dirac:1950pj, Anderson:1951ta, Sundermeyer:1982gv} demanding their stability under dynamical evolution generated by the primary Hamiltonian.
Consistency for the constraints $\phi_{1}$ and $\phi_{3}$ results
in the two secondary constraints\begin{subequations}\label{eq:20}
\begin{gather}
\chi_{1}\equiv\partial_{-}p^{-}+\partial_{i}p^{i}\approx0,\label{eq:20a}\\
\chi_{2}\equiv\partial_{-}\pi^{-}+\partial_{i}\pi^{i}-a^{2}B_{-}\approx0.\label{eq:20b}
\end{gather}
\end{subequations}Eq. (\ref{eq:20a}) reproduces the field equation (\ref{eq:02a}),
while (\ref{eq:20b}) is equivalent to (\ref{eq:02b}) for $\nu=+$,
provided $\phi_{1}=0$ and $\phi_{3}=0$.
For the remaining primary constraints within Eqs (\ref{eq:15}), we have the relations
\begin{subequations}\label{eq:21}
\begin{gather}
\partial_{+}\phi_{2}^{i}\approx\partial_{i}p^{-}+\partial_{k}\left(F^{ki}-a^{2}G^{ki}\right)-2\partial_{-}u_{i}^{2}+2a^{2}\partial_{-}u_{i}^{4}\approx0,\\
\partial_{+}\phi_{4}^{i}\approx\partial_{i}\pi^{-}-a^{2}\left(\partial_{k}F^{ki}+B^{i}\right)+2a^{2}\partial_{-}u_{i}^{2}\approx0,
\end{gather}
\end{subequations}which do not lead to new constraints, but rather result in consistency conditions over the Lagrange multipliers
$u_{i}^{2}$ and $u_{i}^{4}$.

In turn, the set of secondary constraints (\ref{eq:20}) should also be Dirac-Bergmann
consistent. For $\chi_{1}$, the consistency condition is identically
satisfied, but for $\chi_{2}$ it results in the tertiary constraint
\begin{equation}
\chi_{3}\equiv p^{-}+\frac{1}{a^{2}}\pi^{-}-2a^{2}\partial_{-}B_{+}-a^{2}\partial_{i}B^{i}\approx0.\label{eq:22}
\end{equation} 
whose stability requires
$\partial_{+}\chi_{3}\approx0$, leading to
\begin{equation}
\partial_{+}\chi_{3}\approx-\left(1-a^{2}\nabla^{2}\right)B_{+}-2a^{2}\partial_{-}u^{3}\approx0\,.\label{eq:23-1}
\end{equation}
In this last case, once more a condition over a Lagrange multiplier appears and no more
constraints are found from the consistency conditions.  Therefore Eqs. (\ref{eq:15}), (\ref{eq:20}) and (\ref{eq:22}) represent the whole set of constraints for the reduced-order BP model in the null-plane.

\section{Constraints Classification}
At this point, after disclosing the complete set of constraints of the theory, we proceed to compute its algebra in phase space, aiming to understand their roles as gauge symmetry generators and to obtain the proper Dirac brackets needed for canonical quantization -- for that matter, it is imperative to perform the characterization of first-class and second-class constraints.
Since $A_+$ does not show up in the constraint relations and the pair $(A_-,A_i)$ appears only through the combination $F_{-i}$, we identify \begin{subequations}\label{eq:24}
\begin{gather}
\phi_{1}=p^{+}\approx0,\\
\chi_{1}=\partial_{-}p^{-}+\partial_{i}p^{i}\approx0,
\end{gather}
\end{subequations}as a set of first-class constraints.
The remaining constraints can be split into two independent second-class sets.  Indeed, we have
\begin{subequations}\label{eq:25}
\begin{gather}
\phi_{2}^{i}=p^{i}-F_{-i}+a^{2}G_{-i}\approx0,\\
\phi_{4}^{i}=\pi^{i}+a^{2}F_{-i}\approx0,
\end{gather}
\end{subequations}as a first subset of second-class constraints, while 
\begin{subequations}\label{eq:26}
\begin{gather}
\phi_{3}=\pi^{+}\approx0,\\
\chi_{2}=\partial_{-}\pi^{-}+\partial_{i}\pi^{i}-a^{2}B_{-}\approx0,\\
\chi_{3}=p^{-}+\frac{1}{a^{2}}\pi^{-}-2a^{2}\partial_{-}B_{+}-a^{2}\partial_{i}B^{i}\approx0,
\end{gather}
\end{subequations} compose a second subset of second-class constraints.  This can be seen from their PB relations, whose non-null results can be found to be
\begin{subequations}\label{eq:27-1}
\begin{gather}
\left\{ \phi_{2}^{i}\left(x\right),\phi_{2}^{j}\left(y\right)\right\} \approx2\eta^{ij}\partial_{-}^{x}\delta^{3}\left(x-y\right),\\
\left\{ \phi_{2}^{i}\left(x\right),\phi_{4}^{j}\left(y\right)\right\} \approx-2a^{2}\eta^{ij}\partial_{-}^{x}\delta^{3}\left(x-y\right),\\
\left\{ \chi_{2}\left(x\right),\chi_{2}\left(y\right)\right\} \approx2a^{2}\partial_{-}^{x}\delta^{3}\left(x-y\right),\\
\left\{ \chi_{3}\left(x\right),\phi_{3}\left(y\right)\right\} \approx-2a^{2}\partial_{-}^{x}\delta^{3}\left(x-y\right),\\
\left\{ \chi_{3}\left(x\right),\chi_{2}\left(y\right)\right\} \approx\left(1-a^{2}\nabla_{x}^{2}\right)\delta^{3}\left(x-y\right)\,,
\end{gather}
\end{subequations}confirming that (\ref{eq:25}) and (\ref{eq:26}) represent two independent second-class subsets.

From this result, me may build the total Hamiltonian
\begin{equation}
H_{T}\equiv H_{c}+\int_{\Sigma}d\sigma u^{a}\left(x\right)\phi_{a}\left(x\right)+\int_{\Sigma}d\sigma\lambda^{I}\left(x\right)\chi_{I}\left(x\right),\thinspace\thinspace\thinspace\thinspace\thinspace\thinspace I=1,2,3,\label{eq:26-1}
\end{equation}
which adds to the primary Hamiltonian a linear combination of the
secondary constraints. The total Hamiltonian can now be used to
ensure the consistency of the complete set of constraints, resulting
in a coupled set of linear equations in some of the Lagrange multipliers
$\left(u,\lambda\right)$ in (\ref{eq:26-1}). As is well known, consistency
for the first-class constraints does not result in equations for the
respective multipliers. Furthermore, since (\ref{eq:25}) and (\ref{eq:26})
are two independent second-class subsets, the corresponding stability consistency conditions result in two
independent linear equations systems. Usually, the most useful procedure is
to eliminate the second-class constraints with the introduction of
Dirac brackets. But it is sometimes instructive, as is the case
here, to look at the consistency equations themselves.  For the constraints set (\ref{eq:25}), the resulting equations are just the ones already found in
(\ref{eq:21}). For the the second set (\ref{eq:26}), we have\begin{subequations}\label{eq:27}
\begin{align}
\partial_{+}^{x}\chi_{2}\left(x\right) & \approx\chi_{3}\left(x\right)+2a^{2}\partial_{-}^{x}\lambda_{2}\left(x\right)-\left(1-a^{2}\nabla_{x}^{2}\right)\lambda_{3}\left(x\right)\approx0,\label{eq:21a}\\
\partial_{+}^{x}\chi_{3}\left(x\right) & \approx-\left(1-a^{2}\nabla_{x}^{2}\right)B_{+}\left(x\right)+\left(1-a^{2}\nabla_{x}^{2}\right)\lambda_{2}\left(x\right)-2a^{2}\partial_{-}^{x}u_{3}\left(x\right)\approx0,\label{eq:27b}\\
\partial_{+}^{x}\phi_{3}\left(x\right) & \approx\chi_{2}\left(x\right)-2a^{2}\partial_{-}^{x}\lambda_{3}\left(x\right)\approx0.\label{eq:27c}
\end{align}
\end{subequations}
From the last one above, (\ref{eq:27c}), we see that 
$\lambda_{3}$ remounts to an arbitrary function of
$x^{+}$, $x^{1}$ and $x^{2}$ within the constraints surface.  Hence, Eq. (\ref{eq:21a}) leads to
\begin{equation}
\lambda_{2}\left(x\right)\approx\frac{1}{2a^{2}}\left[\partial_{-}^{x}\right]^{-1}\left(1-a^{2}\nabla_{x}^{2}\right)\lambda_{3}\left(x^{+},x^{i}\right),\label{eq:28-1}
\end{equation}
where $\left[\partial_{-}^{x}\right]^{-1}\delta^{3}\left(x-y\right)$
stands for the Green's function of the operator $\partial_{-}^{x}$.

In this case, we may write
\begin{equation}
H_{T}\equiv H_{c}+\int_{\Sigma}d\sigma u^{a}\left(x\right)\phi_{a}\left(x\right)+\int_{\Sigma}d\sigma\lambda_{1}\left(x\right)\chi_{1}\left(x\right)+\int_{\Sigma}d\sigma\rho\left(x\right)\lambda_{3}\left(x^{+},x^{i}\right),\label{eq:29-1}
\end{equation}
where
\begin{eqnarray}
\rho\left(x\right) & \equiv & \chi_{3}\left(x\right)-\frac{1}{2a^{2}}\left[\partial_{-}^{x}\right]^{-1}\left(1-a^{2}\nabla_{x}^{2}\right)\chi_{2}\left(x\right)\nonumber \\
 & = & p^{-}+\frac{1}{2a^{2}}\left(1+a^{2}\nabla^{2}\right)\pi^{-}-2a^{2}\partial_{-}B_{+}-a^{2}\partial_{i}B^{i}\nonumber \\
 &  & -\frac{1}{2a^{2}}\left[\partial_{-}\right]^{-1}\left(1-a^{2}\nabla^{2}\right)\left(\partial_{i}\pi^{i}-a^{2}B_{-}\right).\label{eq:30-1}
\end{eqnarray}
The rewriting of the total Hamiltonian in this form suggests that $\rho\approx0$ could
be chosen as a canonical constraint in place of $\chi_{3}$.  In this way, the constraint
$\chi_{2}$ turns absent from the corresponding linear combination in (\ref{eq:29-1}),
although still present in the canonical Hamiltonian $H_{c}$. Also,
we can see that $\rho(x)$ satisfies
\begin{gather*}
\left\{ \rho\left(x\right),\chi_{2}\left(y\right)\right\} =0,\\
\left\{ \rho\left(x\right),\phi_{3}\left(y\right)\right\} =-2a^{2}\partial_{-}^{x}\delta^{3}\left(x-y\right),\\
\left\{ \rho\left(x\right),\rho\left(y\right)\right\} =\left(1-a^{2}\nabla_{x}^{2}\right)^{2}\left[2a^{2}\partial_{-}^{x}\right]^{-1}\delta^{3}\left(x-y\right),
\end{gather*}
being therefore characterized as second-class.

From this discussion, a redefinition of constraints can certainly ease calculations and enhance clarity.  For that matter, we conveniently redefine and rename the constraint relations as
\begin{subequations}\label{eq:31-1}
\begin{eqnarray}
\Phi_{1} & \equiv & p^{+}\approx0,\label{31-1a}\\
\Phi_{2} & \equiv & \partial_{-}p^{-}+\partial_{i}p^{i},\label{31-1b}\\
\Theta_{1}^{i} & \equiv & p^{i}-F_{-i}+a^{2}G_{-i}\approx0,\label{31-1c}\\
\Theta_{2}^{i} & \equiv & \pi^{i}+a^{2}F_{-i}\approx0,\\
\Gamma_{1} & \equiv & \pi^{+}\approx0,\\
\Gamma_{2} & \equiv & p^{-}+\frac{1}{2a^{2}}\left(1+a^{2}\nabla^{2}\right)\pi^{-}-2a^{2}\partial_{-}B_{+}-a^{2}\partial_{i}B^{i}\nonumber \\
 &  & -\frac{1}{2a^{2}}\left[\partial_{-}\right]^{-1}\left(1-a^{2}\nabla^{2}\right)\left(\partial_{i}\pi^{i}-a^{2}B_{-}\right)\approx0,\\
\chi_{2} & \equiv & \partial_{-}\pi^{-}+\partial_{i}\pi^{i}-a^{2}B_{-}\approx0\label{31-1f}\,,
\end{eqnarray}
\end{subequations}
with (\ref{31-1a})-(\ref{31-1b}) first-class and (\ref{31-1c})-(\ref{31-1f}) second-class.

\section{Dirac brackets}
Since we are working with a highly constrained system, the canonical quantization cannot be achieved by standard Poisson brackets.  Rather, we need Dirac brackets which are tailor-made for constrained systems.  An essential ingredient in that direction is the Dirac matrix and its inverse, constructed from the constraint Poisson brackets.  For the second-class sector, the Dirac matrix can be written as
\begin{equation}
M\left(x,y\right)\equiv\left(\begin{array}{ccc}
A & 0 & 0\\
0 & B & 0\\
0 & 0 & C
\end{array}\right),\label{eq:34}
\end{equation}
with\begin{subequations}\label{eq:35}
\begin{gather}
A\equiv\left\{ \Theta_{a}^{i}\left(x\right),\Theta_{b}^{j}\left(y\right)\right\} =2\eta^{ij}\left(\begin{array}{cc}
1 & -a^{2}\\
-a^{2} & 0
\end{array}\right)\partial_{-}^{x}\delta^{3}\left(x-y\right),\\
B\equiv\left\{ \Gamma_{a}\left(x\right),\Gamma_{b}\left(y\right)\right\} =\left(\begin{array}{cc}
0 & -2a^{2}\partial_{-}^{x}\\
-2a^{2}\partial_{-}^{x} & D_{x}^{2}\left[2a^{2}\partial_{-}^{x}\right]^{-1}
\end{array}\right)\delta^{3}\left(x-y\right),\\
C\equiv\left\{ \chi_{2}\left(x\right),\chi_{2}\left(y\right)\right\} =2a^{2}\partial_{-}^{x}\delta^{3}\left(x-y\right)\,,
\end{gather}
\end{subequations}where $D_{x}\equiv1-a^{2}\nabla_{x}^{2}$.  The block diagonal form of (\ref{eq:34}) clearly shows benefits from the redefinition (\ref{eq:31-1}) as its inverse can be easily obtained by inverting the internal block matrices $A$, $B$ and $C$. Explicitly, we have\begin{subequations}\label{eq:36}
\begin{gather}
A^{-1}=-\eta_{ij}\left(\begin{array}{cc}
0 & 1\\
1 & 1/a^{2}
\end{array}\right)\left[2a^{2}\partial_{-}^{x}\right]^{-1}\delta^{3}\left(x-y\right)+\frac{1}{2}\eta_{ij}\left(\begin{array}{cc}
\alpha_{1} & 0\\
0 & 0
\end{array}\right),\\
B^{-1}=-\left(\begin{array}{cc}
D_{x}^{2}\left[2a^{2}\partial_{-}^{x}\right]^{-2} & 1\\
1 & 0
\end{array}\right)\left[2a^{2}\partial_{-}^{x}\right]^{-1}\delta^{3}\left(x-y\right)-\frac{1}{2a^{2}}\left(\begin{array}{cc}
0 & D_{x}^{2}\left[2a^{2}\partial_{-}^{x}\right]^{-2}\alpha_{2}\\
0 & \alpha_{2}
\end{array}\right)\\
C^{-1}=\left[2a^{2}\partial_{-}^{x}\right]^{-1}\delta^{3}\left(x-y\right),
\end{gather}
\end{subequations}where $\alpha_{1,2}\left(x,y\right)$ are arbitrary symmetric functions of $\left(x^{+},x^{i},y^{+},y^{i}\right)$.  We assume usual boundary conditions such that $\alpha_{1,2}\left(x,y\right)=0$,
indeed necessary to preserve some useful Dirac bracket properties.

From the inverses (\ref{eq:36}), we may readily compute Dirac brackets (DB) defined for two given generic phase space functions $F$ and $G$ as
\begin{eqnarray*}
\left\{ F,G\right\} _{D} & \equiv & \left\{ F,G\right\} -\int d^{3}zd^{3}w\left\{ F,\Theta_{a}^{i}\left(z\right)\right\} \left(A^{-1}\right)_{ij}^{ab}\left(z,w\right)\left\{ \Theta_{b}^{j}\left(w\right),G\right\} \\
 &  & -\int d^{3}zd^{3}w\left\{ F,\Gamma_{a}\left(z\right)\right\} \left(B^{-1}\right)^{ab}\left(z,w\right)\left\{ \Gamma_{b}\left(w\right),G\right\} \\
 &  & -\int d^{3}zd^{3}w\left\{ F,\chi_{2}\left(z\right)\right\} \left(C^{-1}\right)\left(z,w\right)\left\{ \chi_{2}\left(w\right),G\right\} ,
\end{eqnarray*}
which leads to the following fundamental DB relations among the phase space variables:
\begin{eqnarray*}
\left\{ A_{\mu}\left(x\right),A_{\nu}\left(y\right)\right\} _{D} & = & 0,\\
\left\{ A_{\mu}\left(x\right),B_{\nu}\left(y\right)\right\} _{D} & = & -\left(\delta_{\mu}^{i}\delta_{\nu}^{j}\eta_{ij}+\delta_{\mu}^{-}\delta_{\nu}^{+}\right)\left[2a^{2}\partial_{-}^{x}\right]^{-1}\delta^{3}\left(x-y\right),\\
\left\{ A_{\mu}\left(x\right),p^{\nu}\left(y\right)\right\} _{D} & = & \delta_{\mu}^{\nu}\delta^{3}\left(x-y\right)-a^{2}\delta_{\mu}^{j}\left(\delta_{j}^{\nu}\partial_{-}^{x}-\delta_{-}^{\nu}\partial_{j}^{x}\right)\left[2a^{2}\partial_{-}^{x}\right]^{-1}\delta^{3}\left(x-y\right),\\
\left\{ A_{\mu}\left(x\right),\pi^{\nu}\left(y\right)\right\} _{D} & = & 0,\\
\left\{ B_{\mu}\left(x\right),B_{\nu}\left(y\right)\right\} _{D} & = & -\left[\delta_{\mu}^{-}\delta_{\nu}^{-}\partial_{-}^{x}\partial_{-}^{x}+\frac{1}{2a^{2}}\delta_{\mu}^{(-}\delta_{\nu}^{+)}\left(1+a^{2}\nabla_{x}^{2}\right)\right]\left[2a^{2}\partial_{-}^{x}\right]^{-1}\delta^{3}\left(x-y\right)\\
 &  & +\left[
\delta_{\mu}^{(-}\delta_{\nu}^{i)}\partial_{-}^{x}\partial_{i}^{x}+\eta_{\mu i}\left(\delta_{\nu}^{i}+\delta_{\nu}^{j}\partial_{x}^{i}\partial_{j}^{x}\right)\right]\left[2a^{2}\partial_{-}^{x}\right]^{-1}\delta^{3}\left(x-y\right)\\
 &  & -\left[\delta_{\mu}^{+}\delta_{\nu}^{+}D_{x}^{2}\left[2a^{2}\partial_{-}^{x}\right]^{-2}-\delta_{\mu}^{(+}\delta_{\nu}^{i)}D_{x}\partial_{i}^{x}\left[2a^{2}\partial_{-}^{x}\right]^{-1}\right]\left[2a^{2}\partial_{-}^{x}\right]^{-1}\delta^{3}\left(x-y\right),\\
\left\{ B_{\mu}\left(x\right),p^{\nu}\left(y\right)\right\} _{D} & = & a^{2}\eta_{\mu j}\left[\delta_{j}^{\nu}\partial_{-}^{x}-\delta_{-}^{\nu}\partial_{j}^{x}\right]\left[2a^{2}\partial_{-}^{x}\right]^{-1}\delta^{3}\left(x-y\right),\\
\left\{ B_{\mu}\left(x\right),\pi^{\nu}\left(y\right)\right\} _{D} & = & \left(\delta_{\mu}^{\nu}-\delta_{\mu}^{+}\delta_{+}^{\nu}-\frac{1}{2}\delta_{\mu}^{-}\delta_{-}^{\nu}\right)\delta^{3}\left(x-y\right)\\
 &  & -a^{2}\left[\delta_{\mu}^{(+}\eta^{i)\nu}\partial_{i}^{x}-\delta_{\mu}^{+}\delta_{-}^{\nu}D_{x}\left[2a^{2}\partial_{-}^{x}\right]^{-1}\right]\left[2a^{2}\partial_{-}^{x}\right]^{-1}\delta^{3}\left(x-y\right),\\
\left\{ p^{\mu}\left(x\right),p^{\nu}\left(y\right)\right\} _{D} & = & -a^{2}\left(\eta^{\mu j}\delta_{j}^{\nu}\partial_{-}^{x}\partial_{-}^{x}+\delta_{(-}^{\mu}\delta_{j)}^{\nu}\partial_{j}^{x}\partial_{-}^{x}-\delta_{-}^{\mu}\delta_{-}^{\nu}\nabla_{x}^{2}\right)\left[2a^{2}\partial_{-}^{x}\right]^{-1}\delta^{3}\left(x-y\right),\\
\left\{ p^{\mu}\left(x\right),\pi^{\nu}\left(y\right)\right\} _{D} & = & a^{4}\left(\eta^{\mu j}\delta_{j}^{\nu}\partial_{-}^{x}\partial_{-}^{x}+\delta_{(-}^{\mu}\delta_{j)}^{\nu}\partial_{j}^{x}\partial_{-}^{x}-\delta_{-}^{\mu}\delta_{-}^{\nu}\nabla_{x}^{2}\right)\left[2a^{2}\partial_{-}^{x}\right]^{-1}\delta^{3}\left(x-y\right),\\
\left\{ \pi^{\mu}\left(x\right),\pi^{\nu}\left(y\right)\right\} _{D} & = & a^{4}\delta_{-}^{\mu}\delta_{-}^{\nu}\left[2a^{2}\partial_{-}^{x}\right]^{-1}\delta^{3}\left(x-y\right).
\end{eqnarray*}

The Dirac bracket between any second-class constraint and an arbitrary phase space function is identically zero, allowing one to use the
second-class constraints as strong equations simply setting them to zero.  Hence, we have only the first-class constraints (\ref{31-1a})-(\ref{31-1b}) remaining and the total
Hamiltonian becomes
\begin{equation}
H_{T}\equiv H_{c}+\int_{\Sigma}d\sigma u^{r}\left(x\right)\Phi_{r}\left(x\right),\thinspace\thinspace\thinspace\thinspace\thinspace\thinspace\thinspace r=1,2,\label{eq:64}
\end{equation}
with
\begin{equation}
H_{c}=\int_{\Sigma}d\sigma\left[-\frac{1}{2}\frac{1}{a^{4}}\left(\pi^{-}\right)^{2}-\frac{1}{a^{2}}p^{-}\pi^{-}+\frac{1}{4}F_{ij}F^{ij}-\frac{a^{2}}{2}G_{ij}F^{ij}+\frac{a^{2}}{2}B_{i}B^{i}\right],\label{eq:65}
\end{equation}
and\begin{subequations}\label{eq:66}
\begin{eqnarray}
\Phi_{1} & \equiv & p^{+}\approx0,\label{eq:66a}\\
\Phi_{2} & \equiv & \partial_{-}p^{-}+\partial_{i}p^{i}.\label{eq:66b}
\end{eqnarray}
\end{subequations}

We may explicitly check that now we have
\begin{eqnarray*}
\left\{ \Phi_{r}\left(x\right),\Phi_{s}\left(y\right)\right\} _{D} & = & 0,\\
\left\{ \Phi_{r}\left(x\right),H_{c}\right\} _{D} & = & 0,
\end{eqnarray*}
for $r,s=1,2$, characterizing a first-class Abelian dynamical system.  The second-class constraints have been eliminated by means of the constructed algebraic DB phase space structure and the two remaining first-class constraints rightfully signal the gauge freedom of the system, associated to a reduced-order mass generation procedure preserving gauge invariance.

\section{Canonical field equations}
Equipped with the algebraic DB phase space structure, we investigate next the canonical field equations for the $A_\mu$ and $B_\mu$ fields in null-space.  In terms of the natural evolution parameter $\tau=x^{+}$, the main dynamical differential equation for an observable $F$ can be expressed as
\begin{equation}
\partial_{+}F=\left\{ F,H_{T}\right\} _{D}.\label{eq:66-1}
\end{equation}
For the cases $F=A_\pm$, Eq. (\ref{eq:66-1}) results in
\begin{gather}
    \partial_{+}A_{+}\approx u^{1}\\
    \partial_{+}A_{-}\approx-\frac{1}{a^{2}}\pi^{-}-\partial_{-}u^{2}.\label{eq:67}
\end{gather}
signaling an expected gauge freedom captured by the Lagrange multiplier functions.  For the two remaining components of the Podolsky field, considering $F=A_i$ in (\ref{eq:66-1}), we obtain the non-local differential equation
\begin{equation}\label{67-1}
\partial_{+}A_{i}\approx-\left[2a^{2}\partial_{-}\right]^{-1}\left(\partial_{i}\pi^{-}-a^{2}\partial_{j}F_{ji}+a^{2}B_{i}\right)\,.
\end{equation}
Applying the operator $2a^{2}\partial_{-}$ to both sides of (\ref{67-1}) and taking the derivative of (\ref{eq:67}) with respect to $x^i$ into account, we get
\begin{equation}
a^{2}\left(\partial_{\mu}F^{\mu i}+B^{i}\right)\approx a^{2}\partial_{-}\partial^{i}\left(u^{2}+A_{+}\right).\label{eq:68}
\end{equation}
Note that Eq. (\ref{eq:68}) can be made consistent with (\ref{eq:02b}) in the whole phase space by choosing $u^{2}+A_{+}$ to be independent either from $x^{-}$ or $x^i$.  Furthermore, the gauge choice $u^{2}=-A_{+}$, makes both relations
(\ref{eq:68}) and (\ref{eq:67}) in whole phase space equivalent to their corresponding Lagrangian
counterparts.

Concerning the order reducer auxiliary massive field, 
the canonical equation for $B_{+}$ directly obtained from (\ref{eq:66-1}) reads
\begin{equation}
\partial_{+}B_{+}\approx-\left[2a^{2}\partial_{-}\right]^{-2}D\left[\frac{1}{a^{2}}\pi^{-}+p^{-}+a^{2}\partial_{i}B^{i}\right]\,,\label{eq:69}
\end{equation}
while, for $B_{-}$, we have
\begin{equation}
\partial_{+}B_{-}\approx-\frac{1}{2a^{2}}\left(\frac{1}{a^{2}}\pi^{-}+p^{-}+a^{2}\partial_{i}B^{i}\right).
\end{equation}
Finally, for the remaining components $B_{i}$, we get
\begin{equation}\label{canB_}
\partial_{+}B_{i}\approx-\left[2a^{2}\partial_{-}\right]^{-1}\left(\partial_{j}F^{ji}\left(1-a^{4}\right)+a^{2}\partial_{j}\partial^{j}B_{i}+a^{4}\left(\partial_{i}\partial_{j}B^{j}+B_{i}\right)\right).
\end{equation}
We see that non-locality as displayed in the last equations, as well as in the DB structure from where they came, seems to be a price to pay for reducing the derivatives order and eliminating the second-class constraints.

\section{Final considerations}
We have performed the null-plane analysis for a reduced-order version of the Bopp-Podolsky higher-derivative generalized electrodynamics. In the lower-derivatives version, by means of the introduction of an auxiliary vector field, we obtain the action (\ref{eq:01}) containing the Podolsky inverse mass parameter $a$.  In this way, the original Podolsky gauge field is coupled to a massive partner $B_\mu$. After introducing light-front coordinates, we have chosen $x^+$ as a natural parameter for null-plane dynamics and unraveled the theory's constraints structure.  In the null-plane analysis, it has then become clear that $B_\mu$ plays the role of a massive companion field for $A_\mu$ which turns massless as in usual Maxwell theory, corroborating the instant-form results of \cite{Thibes:2016ivt}. Our results are also in line with reference \cite{Bertin:2009gs}, in which the null-plane dynamics of the standard higher-order BP model was investigated. The constraints redefinition (\ref{eq:31-1}) has clearly led to major benefits regarding the separation of the first- and second-class sectors of the theory allowing us to more easily invert the Dirac matrix and compute the Dirac brackets corresponding to the second-class sector.  As a drawback side, we have seen that non-locality seems to abound in some field equations and Dirac bracket expressions due to the appearance of the inverse of a second-order operator proportional to the Podoslky parameter. This is, however, a usual feature of the null-plane Hamiltonian dynamics. Specific physical consequences for the present analysis are currently under investigation, with results to be soon reported by the authors.

\end{document}